\documentclass[review]{elsarticle}
\newcommand{\bbF}{\mathbb{F}}

\newcommand{\F}{\mathbb{F}}

\newcommand{\Q}{\overline{Q}}
\newcommand{\N}{\overline{N}}
\usepackage{mathrsfs}
\usepackage{amssymb}
\usepackage{amsmath}
\usepackage{amsthm}
\usepackage{latexsym}
\usepackage{indentfirst}
\usepackage{enumitem}
\usepackage{anysize}
\usepackage{bbm}

\begin{document}


\title{Two-weight and three-weight codes from trace codes over $\F_p+u\F_p+v\F_p+uv\F_p$ \tnoteref{mytitlenote}}




\author{Yan Liu}
\address{Anhui University, Hefei, Anhui Province 230039, PR China}
\fntext[myfootnote]{The author is supported by NNSF of China (61202068),
Technology Foundation for Selected Overseas Chinese Scholar, Ministry of Personnel of China (05015133), the Open Research Fund of National Mobile Communications Research Laboratory, Southeast University (2015D11) and Key projects of support program for outstanding young talents in Colleges and Universities (gxyqZD2016008).}
\ead{liuyan2612@126.com}

\author[mymainaddress]{Minjia Shi$^{*}$\fnref{myfootnote}}
\ead{smjwcl.good@163.com}

\author[mysecondaryaddress]{Patrick Sol\'e}
\cortext[mycorrespondingauthor]{Corresponding author}
\ead{sole@enst.fr}

\address[mymainaddress]{Key Laboratory of Intelligent Computing $\&$ Signal Processing,
Ministry of Education, Anhui University No. 3 Feixi Road, Hefei Anhui Province 230039, P. R. China, National Mobile Communications Research Laboratory, Southeast University and School of Mathematical Sciences of Anhui University,
Anhui, 230601, P. R. China}
\address[mysecondaryaddress]{CNRS/ LAGA, University of Paris 8, 93 526 Saint-Denis}

%
%
%
%
%
%
%
\begin{abstract}
We construct an infinite family of two-Lee-weight and three-Lee-weight codes over the non-chain ring $\mathbb{F}_p+u\mathbb{F}_p+v\F_p+uv\F_p,$ where $u^2=0,v^2=0,uv=vu.$ These codes are
 defined as trace codes. They have the algebraic structure of abelian codes. Their Lee weight distribution is computed by using Gauss sums.
  With a linear Gray map, we obtain a class of abelian  three-weight codes and two-weight codes over $\F_p$. In particular, the two-weight codes we
  describe are shown to be optimal by application of the Griesmer bound. We also discuss their dual Lee distance. Finally, an application to secret sharing schemes is given.
\end{abstract}

\maketitle


\section{Introduction}
Linear codes with few weights draw motivation from secret sharing \cite{DY2},
combinatorial designs, graph theory \cite{BGH}, association schemes and difference sets \cite{CG,CW}. In addition, they find engineering applications in consumer electronics,
communication and data storage systems. Hence, linear codes with few weights, especially cyclic codes (see \cite{DY}),
have been studied extensively. The determination of their weight distribution leads to difficult arithmetical problems.
 Constacyclic codes over $\F_p+u\F_p+v\F_p+uv\F_p$ have been extensively studied as in \cite{YZK}. This paper is a
generalization of our earlier paper \cite{SLS2,SLS1,SWLP}. Here we consider the codes with few weights over the non-chain ring $R=\F_p+u\F_p+v\F_p+uv\F_p$ with $u^2=v^2=uv-vu=0.$

 It is an interesting problem to construct trace codes. The objective of this paper is to construct the linear codes over $\F_p$ with few weights from the trace codes over an extension ring by using a linear Gray map. These codes turn out to be abelian but possibly not cyclic.
This is noteworthy, as most constructions of codes with few weights in literature
are based on cyclic codes and cyclotomy \cite[\S 9.8.5]{BH}. 
Their weight distribution is determined by using exponential character sums.
After Gray mapping, we obtain an infinite family of $p$-ary abelian codes with few weights.
 In particular, the two-weight codes over $\F_p$ are shown to be optimal for given length and dimension by the application of the Griesmer bound \cite{G}. Furthermore,
an application to secret sharing schemes is sketched out.

The rest of this paper is organized as follows. In Section 2, we define the class of trace codes we are interested in, and present the main results Theorems $1\sim4$ and Proposition 5.
The next section briefly introduces some basic notations and definitions, what is more, we show that trace codes we construct are abliean. Section 3 shows that the code which we constructed and their Gray images are abelian.
 Sections 4 and 5 are devoted to the proof of Theorems $1\sim4$. Section 6 sets up the proof of Proposition 5 and describes an application to secret sharing schemes.
 Section 7 puts the obtained results into perspective, and makes some conjectures for future research.


\section{Statement of main results}
Throughout this paper, let $p$ denote an odd prime. Let ${\mathcal{Q}}$ be the set
of squares in $\F_{p^m}^*$, where $\F_{p^m}^*$ denotes the multiplicative group of nonzero elements of $\F_{p^m}$.
 The set of odd order elements in $\F_{p^m}^*$ is denoted by ${\mathcal{N}}$. Given a positive integer $m>1$, we can construct the ring extension ${\mathcal{R}}=\F_{p^m}+u\F_{p^m}+v\F_{p^m}+uv\F_{p^m}$ of $R=\F_p+u\F_p+v\F_p+uv\F_p$ of degree $m$, where $u^2=0,~v^2=0,~uv=vu.$
  The group of units in ${\mathcal{R}},$ denoted by  ${\mathcal{R}}^*,$ is isomorphic to the direct product $\F_{p}^*\otimes\F_{p}\otimes\F_{p}\otimes\F_{p}.$

  For any $a\in \mathcal{R}$,  the vector $Ev(a)$ is given by the following evaluation map
  $$Ev(a)=(Tr(ax))_{x\in L },$$
  where the definition of $Tr()$ and $L$ are given in next section.
 Under the above map, we define a code $C(m,p)$ by the formula $C(m,p)=\{Ev(a): a\in \mathcal{R}\}.$

 We remark that the definition of this family of linear codes is similar to that \cite{SWLP}. However, here we consider a different base ring.
 The main results of this paper are given below. First, we describe the weight distribution in two Theorems, depending on arithmetical conditions bearing on $m$ and $p.$

 \noindent\textbf{Theorem 1.}\label{enum}
 Assume $m$ is singly-even. Let $\epsilon(p)=(-1)^{\frac{p+1}{2}}.$ For $a\in \mathcal{R}$, the Lee weights of codewords of $C(m,p)$  are as follows.
\begin{enumerate}
\item[(a)] If $a=0$, then $w_L(Ev(a))=0$;
\item[(b)] If $a=\alpha uv \in M\backslash \{0\}$, where $\alpha\in \F_{p^m}^*$, then
\begin{equation*}
w_L(Ev(a))=\begin{cases}
2(p-1)(p^{4m-1}-\epsilon(p)p^{\frac{7m-2}{2}}),~~~\alpha  \in {\mathcal{Q}};\\
2(p-1)(p^{4m-1}+\epsilon(p)p^{\frac{7m-2}{2}}),~~~\alpha \in {\mathcal{N}};
\end{cases}
\end{equation*}
\item[(c)] If $a\in \mathcal{R}\backslash \{\alpha uv : \alpha\in \F_{p^m} \}$, then $$ w_L(Ev(a))= 2(p-1)(p^{4m-1}-p^{3m-1}).$$
 \end{enumerate}
\noindent\textbf{Theorem 2.} Assume $m$ is odd and $p\equiv 3 \pmod{4}.$ For $a\in \mathcal{R}$, the Lee weight of codewords of $C(m,p)$ is given below.
\begin{enumerate}
\item[(a)] If $a=0$, then $w_L(Ev(a))=0$;
\item[(b)] If $a=\alpha uv\in M\backslash \{0\}$, where $\alpha\in \F_{p^m}^*$, then $w_L(Ev(a))=2(p^{4m}-p^{4m-1});$
\item[(c)] If $a\in \mathcal{R}\backslash \{\alpha uv : \alpha\in \F_{p^m} \}$, then $ w_L(Ev(a))= 2(p-1)(p^{4m-1}-p^{3m-1}).$
 \end{enumerate}

 Next, we investigate the dual Lee distance.

 \noindent\textbf{Theorem 3.}
 For all $m> 1,$ the dual Lee distance $d'$ of $C(m,p)$ is $2.$\vspace*{0.2cm}

 Notice that a vector $x$ covers a vector $y$ if $s(x)$ contains $s(y),$ where $s(x)$ and $s(y)$ denotes the support $x$ and $y$, respectively. A \emph{minimal codeword} of a given linear code $C$ over $\F_p$ is a nonzero codeword that does not cover any other nonzero codeword. However, the problem of determining the minimal codewords
of a given linear code is difficult in general.

  Under the linear Gray map which is defined in subsection 3.2, we study the optimality and support structure of the code $\phi(C(m,p))$, in, respectively, Theorem 4 and Proposition 5.

\noindent\textbf{Theorem 4.} Assume $m$ is odd, and $p\equiv 3 \pmod{4}.$ The code $\phi(C(m,p))$ is optimal.

\noindent{\bf Proposition 5.} Under the above theorems, the Gray image $\phi(C(m,p))$ satisfies the following poverties:
\begin{enumerate}
\item[(a)]  All the nonzero codewords of $\phi(C(m,p)),$ for $m$ is even and $m> 2$, are minimal.
\item[(b)] All the nonzero codewords of $\phi(C(m,p)),$ for $m$ is odd and $m\ge 1$, are minimal.
 \end{enumerate}

%
%

\section{Background material}
In this section, we introduce some preliminary results from three parts as follows.\\
\noindent{\bf 3.1.~~Rings and trace function}

There is a Frobenius operator $F()$ which maps $a+bu+cv+duv$ onto $a^p+b^pu+c^pv+d^puv.$  Under the Frobenius operator $F()$, then we can define the following \emph{Trace function}, denoted by $Tr()$,
$$Tr()=\sum_{j=0}^{m-1}F^j().$$
Let $tr()$ be the trace function from $\F_{p^m}$ to $\F_p$, that is, for any $\varepsilon\in \F_{p^m},$ $$tr(\varepsilon)=\varepsilon+\varepsilon^p+\cdots+\varepsilon^{p^{m-1}}.$$
Then it is immediate to check that $$Tr(a+bu+cv+duv)=tr(a)+tr(b)u+tr(c)v+tr(d)uv,$$ for $a,b,c,d \in \F_{p^m},$
and $R$-linearity of $Tr()$ follows from the $\F_p$-linearity of $tr()$.

 In order to be concise, set $M=\{bu+cv+duv:b,c,d\in \F_{p^m}\}$, to denote the unique maximal ideal of $\mathcal{R}.$ The residue field $\mathcal{R}/M$ is isomorphic to $\F_{p^m}$, since $M$ is its maximum ideal.
By direct computation,  we know that ${\mathcal{R}}^*=$ $\{a+bu+cv+duv:a\in \F_{p^m}^{*},b,c,d\in \F_{p^m}\}.$ It is obvious that ${\mathcal{R}}^*$ is not cyclic
and that ${\mathcal{R}}={\mathcal{R}}^*\cup M$. Setting $L=\{a+bu+cv+duv:a\in {\mathcal{Q}},b,c,d\in \F_{p^m}\}$. Thus the set $L$ forms a multiplicative subgroup
of index $2$ of ${\mathcal{R}}^*$. The following proposition introduces a useful property of the trace function $Tr()$.\\
\noindent\textbf{Proposition 6.} If for all $x\in L,$ we have $Tr(ax)=0,$ then $a=0.$
\begin{proof}
Write $x=x_0+x_1u+x_2v+x_3uv$ and $a=a_0+a_1u+a_2v+a_3uv,$ with $x_0\in\mathcal{Q}, x_i,a_j\in \F_{p^m},~i=1,2,3,j=0,1,2,3.$ Thus $ax=a_0x_0+(a_0x_1+a_1x_0)u+(a_0x_2+a_2x_0)v+(a_0x_3+a_1x_2+a_2x_1+a_3x_0)uv$ and $Tr(ax)=0$
is equivalent to a system of equations, that is, $tr(a_0x_0)=0,tr(a_0x_1+a_1x_0)=0,tr(a_0x_2+a_2x_0)=0$ and $tr(a_0x_3+a_1x_2+a_2x_1+a_3x_0)=0.$ Thus, we are interested in this system of equations with indeterminate $a_i,i=0,1,2,3$. From the nondegenerate character of $tr()$ \cite{MS}, we get $a=0$ by solving these equations.
\end{proof}

\noindent{\bf 3.2.~~Codes and Gray map}

  A {\bf linear code} $C$ over $R$ of length $n$ is an $R$-submodule of $R^n$. A {\bf codeword} is any element of the code $C$. For any $x=(x_1,x_2,\dots,x_n), y=(y_1,y_2,\dots,y_n)\in R^n$, their standard inner product
  is defined by $\langle x,y\rangle=\sum_{i=1}^nx_iy_i$, where the operation is performed in $R$. The {\bf dual code} of $C$, denoted by $C^\perp$, consists of all vectors of $R^n$ which are orthogonal to every codeword in $C$, i.e.,  $C^\perp=\{y\in R^n|\langle x,y\rangle =0, \forall x\in C\}.$

  We note that the Lee weight of an element $a+bu+cv+duv\in R$ was defined in \cite{YZK} to
be the Hamming weight of the $p$-ary vector $(d,c+d,b+d,a+b+c+d)$. This leads to
the Gray map $\phi:~R^n\rightarrow\F_p^{4n}$:
$$\phi(\mathbf{a}+\mathbf{b}u+\mathbf{c}v+\mathbf{d}uv)=(\mathbf{d},\mathbf{c}+\mathbf{d},\mathbf{b}+\mathbf{d},
\mathbf{a}+\mathbf{b}+\mathbf{c}+\mathbf{d} ), $$
where $\mathbf{a},\mathbf{b},\mathbf{c},\mathbf{d} \in \F_p^n.$
 As was observed in \cite{YZK}, $\phi$ is a distance preserving isometry from $(R^n,d_L)$ to $(\F_p^{4n},d_H)$, where $d_L$ and $d_H$ denote the Lee and Hamming distance in $R^n$ and $\F_p^{4n}$, respectively. That means if $C$ is a linear code over $R$ with parameters $(n,p^k,d)$, then $\phi(C)$ is a linear code of parameters $[4n,k,d]$ over $\F_p$.

 Given a finite abelian group $G,$ a code over $R$ is said to be {\bf abelian} if it is an ideal of the group ring $R[G].$ In other words, the coordinates of $C$ are indexed by elements of $G$ and $G$ acts regularly on this set. In the special case when $G$ is cyclic, the code is a cyclic code in the usual sense \cite{MS}. Hence, we have the following proposition.

\noindent\textbf{Proposition 7.} The group $L$ acts regularly on the coordinates of $C(m,p).$
\begin{proof}
For any $v',u' \in L$, the change of variables $ x\mapsto (u'/v')x$ permutes the coordinates of $C(m,p),$ and maps $v'$ to $u'.$ This defines a transitive action on the coordinates of $C(m,p).$
Such a permutation is unique, given $v',u'.$ Hence the action is regular.
\end{proof}
The code $C(m,p)$ is thus an {\em abelian code} with respect to the group $L$ from Proposition 7. It means that $C(m,p)$ is an ideal of the group ring $R[L].$
As observed in the previous subsection, $L$ is a not cyclic group, hence $C(m,p)$ may be not cyclic. The next result shows that the Gray image of $C(m,p)$ is also abelian.\\
\noindent\textbf{Proposition 8.} A finite group of size $4|L|$ acts regularly on the coordinates of $\phi(C(m,p)).$
 \begin{proof}
By the definition of Gray map, $\phi(\mathbf{a}+\mathbf{b}u+\mathbf{c}v+\mathbf{d}uv)=(\mathbf{d},\mathbf{c}+\mathbf{d},\mathbf{b}+\mathbf{d},
\mathbf{a}+\mathbf{b}+\mathbf{c}+\mathbf{d}),$ where $\mathbf{a},\mathbf{b},\mathbf{c},\mathbf{d}\in \F_p^n$.
Now if $\mathbf{a}+\mathbf{b}u+\mathbf{c}v+\mathbf{d}uv \in C(m,p),$ then by linearity,
$$(1+u)(\mathbf{a}+\mathbf{b}u+\mathbf{c}v+\mathbf{d}uv)=\mathbf{a}+(\mathbf{a}+\mathbf{b})u
+\mathbf{c}v+(\mathbf{c}+\mathbf{d})uv \in C(m,p)$$, $$(1+v)(\mathbf{a}+\mathbf{b}u+\mathbf{c}v+\mathbf{d}uv)=\mathbf{a}+\mathbf{b}u
+(\mathbf{a}+\mathbf{c})v+(\mathbf{b}+\mathbf{d})uv \in C(m,p)$$ and
$(1+u+v+uv)(\mathbf{a}+\mathbf{b}u+\mathbf{c}v+\mathbf{d}uv)=\mathbf{a}+(\mathbf{a}+\mathbf{b})u
+(\mathbf{a}+\mathbf{c})v+(\mathbf{a}+\mathbf{b}+\mathbf{c}+\mathbf{d}) uv\in C(m,p)$.
Further $\phi(\mathbf{a}+(\mathbf{a}+\mathbf{b})u
+\mathbf{c}v+(\mathbf{c}+\mathbf{d})uv)=(\mathbf{c}+\mathbf{d},\mathbf{d},
\mathbf{a}+\mathbf{b}+\mathbf{c}+\mathbf{d},\mathbf{b}+\mathbf{d})$, $\phi(\mathbf{a}+\mathbf{b}u
+(\mathbf{a}+\mathbf{c})v+(\mathbf{b}+\mathbf{d})uv)=(\mathbf{b}+\mathbf{d},
\mathbf{a}+\mathbf{b}+\mathbf{c}+\mathbf{d},\mathbf{d},\mathbf{c}+\mathbf{d})$, and
$\phi(\mathbf{a}+(\mathbf{a}+\mathbf{b})u
+(\mathbf{a}+\mathbf{c})v+(\mathbf{a}+\mathbf{b}+\mathbf{c}+\mathbf{d})uv)
=(\mathbf{a}+\mathbf{b}+\mathbf{c}+\mathbf{d},\mathbf{b}+\mathbf{d},\mathbf{c}+\mathbf{d},\mathbf{d})$,
so that $\phi(C(m,p))$ is invariant under an involution that permutes the four parts of a codeword. Thus, $\phi(C(m,p))$ is invariant under the regular action of
a group of order $4 \vert L\vert.$
\end{proof}

 \noindent{\bf3.3.~~Character sums}

Let $\psi$  denote the canonical additive character of $\F_{p^m}.$
Let $\chi$ be an arbitrary multiplicative character of $\F_{p^m}.$
The quadratic multiplicative character of $\F_{p^m}$ is denoted by $\eta$ which is defined by
\begin{equation*}
 \eta(x)=\begin{cases}
 1, ~~~~~\mathrm{if}~x\in{\mathcal{Q}}; \\
 -1,~~~\mathrm{if}~x\in {\mathcal{N}}.
  \end{cases}
\end{equation*}
The classical {\bf Gauss sum} attached to $\chi$ can be defined as $$G(\chi)=\sum_{x\in \F_{p^m}^*}\psi(x)\chi(x).$$
It is well-known that the explicit evaluation of a Gauss sum is a very difficult problem for a general $\chi$.
But when the order of $\chi$ equals two, then the explicit values of corresponding Gauss sum are known and are recorded in \cite[Theorem 5.15]{LN} as follows:
\begin{equation}\label{im}
G(\eta)=\begin{cases}
 (-1)^{m-1}p^{\frac{m}{2}},~~~~p\equiv 1~(\mathrm{mod}~4); \\
 (-1)^{m-1}i^mp^{\frac{m}{2}}, ~p\equiv 3~(\mathrm{mod}~4).
  \end{cases}
\end{equation}
In particular, in the case when $m$ is singly-even,
we know that $G(\eta)=\epsilon(p)p^{\frac{m}{2}},$ with $\epsilon(p)=(-1)^{\frac{(p+1)}{2}}.$


We now study two character sums $\sum\limits_{x\in {\mathcal{Q}}}\psi(x)$ and $\sum\limits_{x\in {\mathcal{N}}}\psi(x)$, denoted by $\Q$ and $\N$, respectively.
It is immediate from the orthogonal property of additive characters which can be found in \cite[Lemma 9, p. 143]{MS} that $\Q+\N=-1.$
 Further, it can be shown that $\Q=\frac{\epsilon(p)p^{\frac{m}{2}}-1}{2}$ and $\N=-\frac{\epsilon(p)p^{\frac{m}{2}}+1}{2}$ by observing that the characteristic function of ${\mathcal{Q}}$ is $\frac{1+\eta}{2}.$
{\bf\section{ Proof of Theorems 1, 2 and 4}}
 This section is divided into four parts. First, the statement of some lemmas. Next, the proof of Theorems 1, 2 and 4. For the rest of this paper, for convenience, we adopt the following notations unless stated otherwise
in this paper. Let $\omega=\exp(\frac{2\pi i}{p})$ and $N=4|L|=2(p^{4m}-p^{3m})$.\\
\noindent{ \bf 4.1.~~Statement of some lemmas}

If $y=(y_1,y_2,\dots,y_N)\in \mathbb{F}_p^N,$ let $$\Theta(y)=\sum_{j=1}^N\omega^{y_j}.$$
For simplicity, we let $\theta(a)=\Theta(\phi(Ev(a))).$  Taking account of the linear property of the Gray map and  the evaluation map, we see that $\theta(sa)=\Theta(\phi(Ev(sa))),$ for any $s\in \F_p^*.$ Now, we present the following lemmas, which play an important role in the process of proof in Theorems 1, 2 and 4.

%
 \noindent\textbf{Lemma 9.} \cite[Griesmer bound]{G,MS} If $[N,K,d]$ are the parameters of a linear $p$-ary code, then
$$\sum_{j=0}^{K-1}\Big\lceil \frac{d}{p^j} \Big\rceil \le N.$$
\noindent\textbf{Lemma 10.}~\cite[Lemma 1]{SWLP}\label{5.1} For all $y=(y_1,y_2,\dots,y_N)\in \mathbb{F}_p^N,$ we have
$$\sum_{s=1}^{p-1}\Theta(sy)=(p-1)N-pw_H(y).$$\\
\noindent\textbf{Lemma 11.}~\cite[Lemma 4.2]{SLS1} If $z \in \mathbb{F}_{p^m}^*,$ then $$\sum\limits_{x\in \mathbb{F}_{p^m}}\omega^{tr(z x)}=0.$$
\noindent\textbf{Lemma 12.}~\cite[Lemma 2]{SWLP} Set $\Re$ denotes the real part of complex number. If $p\equiv 3 \pmod{4},$ then $$\sum_{s=1}^{p-1}\theta(sa)=(p-1)\Re(\theta(a)).$$

Combining Lemma 10 and the definition of Gray map, for $Ev(a)\in C(m,p)$, we have
\begin{eqnarray}
 w_L(Ev(a) ) &=& \frac{(p-1)N- \sum\limits_{s=1}^{p-1}\Theta(s\phi(Ev(a)))}{p} \nonumber  \\
 &=& \frac{(p-1)N-\sum\limits_{s=1}^{p-1} \theta(sa)}{p} .
\end{eqnarray}
When $m$ is odd and $p\equiv 3 \pmod{4}$, from Equation (2) and Lemma 12, we deduce that
 \begin{equation}\label{3}
   w_L(Ev(a))=\frac{p-1}{p}(N-\Re(\theta(a))).
 \end{equation}
In the light of Equations (2) and (3), it is easy to see that the value of $\theta(a)$ is the key points in the computation of Lee weight of the codeword $Ev(a)$, where $a\in \mathcal{R}$.\\

\noindent{\bf 4.2.~~Proof of Theorem 1}\vspace*{0.2cm}

\begin{proof}
Setting $x=x_0+x_1u+x_2v+x_3uv,$ where $x_0\in \mathcal{Q},x_1,x_2,x_3\in \F_{p^m}$ throughout the process of proof in this theorem. Let $a=a_0+a_1u+a_2v+a_3uv\in \mathcal{R}$, where $a_i \in \F_{p^m},~i=0,1,2,3$, then we get $$ax=a_0x_0+(a_0x_1+a_1x_0)u+(a_0x_2+a_2x_0)v+(a_0x_3+a_1x_2+a_2x_1+a_3x_0)uv, $$ and
\begin{eqnarray*}
 Tr(ax) &=&tr(a_0x_3+a_1x_2+a_2x_1+a_3x_0)uv+tr(a_0x_2+a_2 x_0)v+tr(a_0x_1+a_1x_0)u+tr(a_0x_0)\\
 &=:&D_3uv+D_2v+D_1u+D_0.
\end{eqnarray*}
Taking Gray map yields
\begin{eqnarray}
  \phi(Ev(a)) &=& (D_3,D_2+D_3,D_1+D_3,D_0+D_1+D_2+D_3)_x  \nonumber \\  \nonumber
   &=& (tr(a_0x_3+a_1x_2+a_2x_1+a_3x_0),tr(a_0x_2+a_2 x_0+a_0x_3+a_1x_2+a_2x_1+a_3x_0), \nonumber \\
 &&tr(a_0x_1+a_1x_0+a_0x_3+a_1x_2+a_2x_1+a_3x_0),tr(a_0x_0+a_0x_1+a_1x_0+a_0x_2+a_2 x_0 \nonumber \\
 &&+a_0x_3+a_1x_2+a_2x_1+a_3x_0) )_{x_0,x_1,x_2,x_3}.
\end{eqnarray}
Taking character sum\ \
\begin{eqnarray*}
  \theta(a) &=& \sum_{i=0}^3\sum_{x_1,x_2,x_3\in \F_{p^m}}\sum_{x_0\in \mathcal{Q}}\omega^{D_i} .
\end{eqnarray*}
Assume $\F_{p^m}^*=<\xi>,$ we get then $\F_p^*=<\xi^{\frac{p^m-1}{p-1}}>$. Note that $2|\frac{p^m-1}{p-1}$ because of $gcd(2,m)=2.$ Thus we claim that $s\in \F_p^*$ is a square in $\F_{p^m}$, which implies $ \theta(sa)= \theta(a)$. From Equation (2), we have
\begin{eqnarray}
 w_L(Ev(a) ) &=& \frac{(p-1)N-\sum\limits_{s=1}^{p-1} \theta(sa)}{p} =\frac{(p-1)N-(p-1)\theta(a)}{p}.
\end{eqnarray}

(a) If $a=0$, then $Ev(a)=(\underbrace{0,0,\cdots,0}\limits_{|L|})$. So $w_L(Ev(a))=0$.\\

(b) Let $a=\alpha uv,$ where $\alpha\in \F_{p^m}^{*}$. Taking $a_0=a_1=a_2=0,a_3=\alpha$ in Equation (4), then we can easily get
\begin{eqnarray*}
  \theta(a) &=& 4\sum_{x_1,x_2,x_3\in \F_{p^m}}\sum_{x_0\in \mathcal{Q}}\omega^{tr(\alpha x_0)} = 4p^{3m}\sum_{x_0\in \mathcal{Q}}\omega^{tr(\alpha x_0)}\\
   &=&\begin{cases}
   4p^{3m}\overline{Q},~~~~~~~~~\alpha \in \mathcal{Q};\\
   4p^{3m}\overline{N},~~~~~~~~~\alpha \in \mathcal{N}.
   \end{cases}
\end{eqnarray*}
 By Equation (4), we have
\begin{eqnarray*}
 w_L(Ev(a) ) &=& \begin{cases}
   \frac{(p-1)}{p}(N-4p^{3m}\overline{Q}),~~~\alpha \in \mathcal{Q};\\
     \frac{(p-1)}{p}(N-4p^{3m}\overline{N}),~~~\alpha \in \mathcal{N}.
   \end{cases}
\end{eqnarray*}\vspace*{0.1cm}

(c) When $a\in \mathcal{R}\backslash \{\alpha uv : \alpha\in \F_{p^m} \}$, we obtain $\theta(a)=0$ by using a similar approach in the case (b). Hence, we deduce  $ w_L(Ev(a) )= \frac{(p-1)N}{p}$ from Equation (5).

 %
%
%
\end{proof}\vspace*{0.2cm}

Armed with Theorem 1 and Proposition 6, we have constructed a class of $p$-ary linear code of length $N=2p^{4m}-2p^{3m},$ dimension $4m,$ with three nonzero weights $w_1<w_2<w_3$ of values
\begin{eqnarray*}
w_1&=&2(p-1)(p^{4m-1}-p^{\frac{7m-2}{2}}),\\
w_2&=&2(p-1)(p^{4m-1}-p^{3m-1}),\\
w_3&=&2(p-1)(p^{4m-1}+p^{\frac{7m-2}{2}}),
\end{eqnarray*}
with respective frequencies $f_1,f_2,f_3$ given by
\begin{eqnarray*}
f_1&=&\frac{p^m-1}{2},~~~f_2=p^{4m}-p^{m},~~~f_3=\frac{p^m-1}{2}.
\end{eqnarray*}
(Note that taking $\epsilon(p)=1,$ or $-1,$ leads to the same values of $w_1$ and $w_3.$)

\noindent{\bf Example 13.}
 Let $p=3$ and $m=2.$ We obtain a ternary code of parameters $[11664,8,5832].$ The nonzero weights are 5832, 7776 and 11664, of frequencies 4, 6552 and 4, respectively.

\noindent {\bf4.3.~~Proof of Theorem 2}

\begin{proof}
 The cases (a) and (c) are like in the proof of Theorem 1. Then it suffices to prove the case (b). From the process of proof (b) in Theorem 1, it is easy to know that $\Re(\theta(a))=-2p^{3m}$. Thus, we get $w_L(Ev(a))=2(p^{4m}-p^{4m-1})$ from Equation (3). This completes the proof of Theorem 2.
\end{proof}

In view of Theorem 2 and Proposition 6, we obtain a class of $p$-ary two-weight linear codes of parameters $[2p^{4m}-2p^{3m},4m],$ with two nonzero weights $w_1<w_2$ given by
\begin{eqnarray*}
w_1&=&2(p-1)(p^{4m-1}-p^{3m-1}),\\
w_2&=&2(p^{4m}-p^{4m-1}),
\end{eqnarray*}
with respective frequencies $f_1,f_2$ given by
\begin{eqnarray*}
f_1&=&p^m-1,~~~f_2=p^{4m}-p^{m}.
\end{eqnarray*}

\noindent{\bf Example 14.}
 Let $p=3$ and $m=1.$ We obtain a ternary code of parameters $[108,4,72].$ The nonzero weights are 72 and 108, of frequencies 2 and 78, respectively. Note that this code is an optimal ternary code.\\
\noindent{\bf4.3.~~Proof of Theorem 4 }
\begin{proof}
We apply Lemma 9, that is to say the Griesmer
bound, with
 $N=2p^{4m}-2p^{3m},\, K=4m,$ and $d=2(p-1)(p^{4m-1}-p^{3m-1}).$ Then, there exist three possible values for the ceiling function, depending on the position of $j,$ as follows:
\begin{itemize}
 \item $0\leq j\le 3m-1 \Rightarrow \lceil \frac{d+1}{p^j} \rceil =2(p-1)(p^{4m-j-1}-p^{3m-j-1})+1,$
  \item $j= 3m \Rightarrow \lceil \frac{d+1}{p^j} \rceil =2(p-1)p^{m-1}-1,$
 \item $3m<j\leq 4m-1 \Rightarrow \lceil \frac{d+1}{p^j} \rceil =2(p-1)p^{4m-j-1}.$
\end{itemize}
Thus \begin{eqnarray*}
       \sum_{j=0}^{K-1}\Big\lceil \frac{d+1}{p^j} \Big\rceil&=& \sum_{j=0}^{3m-1}\Big\lceil \frac{d+1}{p^j}\Big\rceil+\sum_{j=3m+1}^{4m-1}\Big\lceil \frac{d+1}{p^j}\Big\rceil +\Big\lceil \frac{d+1}{p^{3m}}\Big\rceil \\
        &=& 2(p-1)\sum_{j=0}^{3m-1}(p^{4m-j-1}-p^{3m-j-1})+3m+2(p-1)\sum_{j=3m+1}^{4m-1}p^{4m-j-1} \\
        &&+2(p-1)p^{m-1}-1\\
        &=&2p^{4m}-2p^{3m}+3m-1.
     \end{eqnarray*}
     Hence, Theorem 3 is proved by observing that $ \sum_{j=0}^{K-1}\lceil \frac{d+1}{p^j} \rceil-N=2p^{4m}-2p^{3m}+3m-1-(2p^{4m}-2p^{3m})=3m-1>0.$
\end{proof}
\section{Proof of Theorem 3}
We investigate the dual Lee distance of $C(m,p)$ in this section. As before we study another property of the trace function. The proof of the following lemma is similar to Proposition 6, so we omit it here.\\
\noindent{\bf Lemma 15.}
 If for all $a \in \mathcal{R},$ we have $Tr(ax)=0,$ then $x=0.$

%
 Armed with Lemma 15 and sphere-packing bound, we can prove Theorem 3 as follows: first, we show that $d'<3.$ If not, we can apply the sphere-packing bound to $\phi(C(m,p)^\bot),$ to obtain
 $$p^{4m}\ge 1+N(p-1)=1+2(p^{4m}-2p^{3m})(p-1),$$ or, after expansion
 $$3p^{4m}-2p^{3m}\ge 1+ 2(p^{4m+1}-p^{3m+1}). $$
 Dropping the $1$ in the RHS, and then we deduce $3p^{m}-2>2(p^{m+1}-p)$ from
 dividing both sides by $p^{3m}.$ Since $p>2$ is an odd prime and $m\geq1$,
 we can write, however
 \begin{eqnarray*}
   3p^{m}-2-2(p^{m+1}-p) &=& p^m(3-2p)+2p-2 = (3-2p)(p^m-1)+1 \\
    &<& 3-2p+1\\
    &=&2(2-p)<0,
 \end{eqnarray*}
  a contracdiction.

 Next, we check that $d'= 2,$ by showing that $C(m,p)^\bot,$ does not contains a codewords of Lee weight one.
 If it does, let us assume first that it has value $\gamma$ at some $x \in \mathcal{R}^*.$ Thus $\gamma\in \{1,-1+u,-1+v,1-u-v+uv\}$, a subset of units. This implies that $\forall a \in \mathcal{R},\gamma
 Tr(ax)=0,$ then we must have $Tr(ax)=0$ since $\gamma$ is a unit, and by using Lemma 15, we know $x=0.$ Contradiction with $x\in L.$ So $d'=2.$

\section{Proof of Proposition 5 and secret sharing schemes}

In general determining the minimal codewords, which is defined in section 2
of a given linear code is a difficult task. However, if the weights of a given linear code $C$ over $\F_p$ are close enough to each other, then each nonzero codeword of $C$ is minimal, as described by the following lemma \cite{AB}.

\noindent{\bf Lemma 16.} \cite{AB} Denote by $\omega_0$ and $\omega_{\infty}$ the minimum and maximum nonzero weights of a given $p$-ary linear code $C$, respectively. If
$$\frac{\omega_0}{\omega_{\infty}}>\frac{p-1}{p},$$ then every nonzero codeword of $C$ is minimal.\ \

By the above lemma, Proposition 5 is proved as follows: rewriting the inequality of  Lemma 16 as $p\omega_0>(p-1)\omega_{\infty}$. First, we consider the case $m>2$ and $m$ is even.
Deduce from $p\omega_0>(p-1)\omega_{\infty}$ we have $p^{\frac{m}{2}}>2p-1$, with $\omega_0=2(p-1)(p^{4m-1}-p^{\frac{7m-2}{2}})$ and $\omega_{\infty}=2(p-1)(p^{4m-1}+p^{\frac{7m-2}{2}}).$
This completes the proof of (a) of Proposition 5 because of $m>2.$  Now, we deal with the case that $m$ is odd and $p\equiv 3~(\mathrm{mod}~4)$. Using the criterion in the previous inequality with $\omega_0=2(p-1)(p^{4m-1}-p^{3m-1})$ and $\omega_{\infty}=2(p^{4m}-p^{4m-1})$, dividing both sides by $2(p-1)$,
we end up with the condition $$p(p^{4m-1}-p^{3m-1})>p^{4m}-p^{4m-1},$$
 or $ p^{4m-1}-p^{3m-1}>0$, which is true for $m\geq1.$ Then a routine computation completes the proof of (b).\vspace*{0.1cm}
%
%

  To determine the set of all minimal access sets of a secret sharing scheme, the notion of minimal codewords was introduced. The Massey's scheme is a construction of a secret sharing scheme (SSS) using a code $C$ of length $N$ over $\F_p.$
It is well-known that linear codes have application in SSS \cite{YD}. On the other hand, it is worth mentioning that in some special cases, that is, when all nonzero codewords are minimal, it was shown in \cite{DY2} that there is the following alternative, depending on $d'$:

\begin{itemize}
 \item If $d'\ge 3,$ then the SSS is \emph{``democratic''}: every user belongs to the same number of coalitions,
 \item If $d'=2,$  then there are users  who belong to every coalition: the \emph{``dictators''}.
\end{itemize}
One or the other situation might be more suitable depending on the application.
In view of Proposition 5 and Theorem 3, we see that a SSS built on $\phi(C(m,p))$ is dictatorial.
\section{Conclusion}
In the present work, we have studied a family of trace codes over the ring $\bbF_p+u\bbF_p+v\F_p+uv\F_p,$
with $u^2=0, v^2=0, uv=vu.$ These codes are provably abelian, but not visibly cyclic.
We have been able to employ their Lee weight distribution relying on classical Gauss sums, and yielding a family of abelian $p$-ary two-weight codes by using a linear Gray map.
 The two-weight codes are shown to be optimal by use of the Griesmer bound. It is worth exploring more general constructions by varying the alphabet of the code, the Gray map,
 or the localizing set of the trace code.

\section*{References}

\bibliography{mybibfile}

\end{document}